\documentclass[twoside,11pt]{article}

\usepackage{blindtext}

\usepackage[abbrvbib,preprint,nohyperref]{jmlr2e}

\usepackage{svg}
\usepackage{amsmath}
\usepackage{graphicx}
\usepackage{tabularx}
\usepackage{listings}
\usepackage[shortlabels]{enumitem}
\usepackage{subcaption}
\usepackage{booktabs}
\usepackage{hyperref}
\usepackage{algorithm}
\usepackage{graphics}
\usepackage{algpseudocode}

\usepackage{tikz}
\usetikzlibrary{calc,angles,positioning,intersections,quotes,decorations.markings}
\usepackage{tkz-euclide}
\usepackage{float}

\def\bx{\mathbf{x}}

\def\bX{\mathbf{X}}
\def\bY{\mathbf{Y}}

\def\bU{\mathbf{U}}
\def\bV{\mathbf{V}}

\def\b0{\mathbf{0}}
\def\b1{\mathbf{1}}

\def\bmu{\mbox{\boldmath $\mu$}}

\def\bbeta{\mbox{\boldmath $\beta$}}

\def\btheta{\mbox{\boldmath $\theta$}}

\def\bSigma{\mbox{\boldmath $\Sigma$}}

\def\bLambda{\mbox{\boldmath $\Lambda$}}

\def\bTheta{\mbox{\boldmath $\Theta$}}

\DeclareMathOperator*{\argmax}{arg\,max}

\newcommand{\matsig}{\mathbf{\Sigma}}

\usepackage{lastpage}
\jmlrheading{23}{2022}{1-\pageref{LastPage}}{1/21; Revised 5/22}{9/22}{21-0000}{Author One and Author Two}

\ShortHeadings{Mixture Models: Python Library}{Kasa et. al.}
\firstpageno{1}

\begin{document}

\title{\texttt{Mixture-Models}: a one-stop Python Library for Model-based Clustering using various Mixture Models}

\author{\name Siva Rajesh Kasa\thanks{equal contribution} \thanks{This work does not relate to the author's position at Amazon}\email kasa@u.nus.edu
       \AND
       \name Hu Yijie\footnotemark[1] \email e0925542@u.nus.edu 
        \AND
        \name Santhosh Kumar Kasa\footnotemark[1] \footnotemark[2] \email mixturemodelscollab@gmail.com 
        \AND
       \name Vaibhav Rajan \email vaibhav.rajan@nus.edu.sg\\
       \addr School of Computing\\
       National University of Singapore\\
       Singapore, 117417}

\editor{xxxxx}

\maketitle

\begin{abstract}%

\texttt{Mixture-Models} is an open-source Python library for fitting Gaussian Mixture Models (GMM) and their variants, such as Parsimonious GMMs, Mixture of Factor Analyzers, MClust models, Mixture of Student's t distributions, etc. It streamlines the implementation and analysis of these models using various first/second order optimization routines such as Gradient Descent and Newton-CG through automatic differentiation (AD) tools. This helps in extending these models to high-dimensional data, which is first of its kind among Python libraries.  The library provides user-friendly model evaluation tools, such as BIC, AIC, and log-likelihood estimation. The source-code is licensed under MIT license and can be accessed at \url{https://github.com/kasakh/Mixture-Models}.
The package is highly extensible, allowing users to incorporate new distributions and optimization techniques with ease. We conduct a large scale simulation to compare the performance of various gradient based approaches against Expectation Maximization on a wide range of settings and identify the corresponding best suited approach.
\end{abstract}

\begin{keywords}
  Mixture Models, Automatic Differentiation, Clustering
\end{keywords}

\section{Introduction}

Gaussian Mixture Models (GMM) are one of the widely studied model-based clustering approaches.
There have been several variants of GMMs proposed such as Mixture of Factor Analyzers (MFA) \citep{Ghah:Hilt:1997}, Parsimonious GMMs (PGMM) \citep{McNi:Murp:Pars:2008},
Student's \textit{t} Mixture Models (TMM) \citep{peel2000robust}, MClust \citep{scrucca2016mclust}, etc. which bake in additional modeling assumptions and constraints. These variants offer several advantages over fitting a unconstrained vanilla GMM model by addressing one or many of the following aspects.
\begin{itemize}
    \item \textbf{Dimensionality reduction}: Imposing additional constraints or underlying structure on top of GMM can mitigate the curse of dimensionality by reducing the number of parameters to be estimated, making the model suitable for high dimensional data.

    \item \textbf{Computational efficiency}: Fewer parameters can lead to faster computation and convergence during model fitting. This efficiency is especially beneficial for large datasets.

    \item \textbf{Robustness}: Imposing structure and reducing the overparametrization can make the model more robust to initialization and outliers. 
    Plain GMMs have degenerate maximum likelihood and are hence sensitive to initialization and outliers \citep{day1969estimating}; some of these variants don't suffer from this degeneracy issue.

    \item \textbf{Interpretability}: Variants which can account for latent factors (MFAs, PGMMs, etc.) can capture the underlying structure of the data, making it easier to interpret the relationships between variables. 
    
\end{itemize}

There are several R packages such as \citet{pgmmrpackage} and \citet{mclustrpackage} which can fit these variants using Expectation Maximization (EM) \cite{dempster1977maximum}. However when it comes to modeling high dimensional data (where the number of free parameters $\gtrapprox$ the number of datapoints $n$), there is a two-fold problem with the above mentioned R packages. Firstly, the use of EM based inference makes some of the models in these packages not suitable for HD data as in the maximization (M) step the covariance matrix estimate is not full rank.  Second, to address this full rank issue these packages only recommend to fit highly constrained models. However, imposing such hard modeling constraints may be too stringent and can %
	degrade clustering performance as they do not take into account 
	feature dependencies that can vary across the clusters, and may lead to incorrect orientations or shapes of the inferred clusters.
	As a result, pre-determined constraints on means or covariance matrices are discouraged and a more data-dependent discovery of cluster structure is recommended, e.g., by \citep{zhou2009penalized,fop2019model}. 

 To tackle this full rank issue associated with EM and enable fitting of minimally constrained models on HD data, a gradient-based (GD) inference through the use of Automatic Differentiation (AD) tools and certain reparametrization techniques are proposed \citep{kasa2020gaussian,kasa2023avoiding}. Following this line of thought, we chose to develop the \texttt{Mixture-Models} library in Python, primarily due to the availability of a pre-existing package \texttt{autograd}~\citep{maclaurin2015autograd} implementing AD for customizeable objective functions. Apart from the above mentioned issue of HD data, using AD tools eliminate the need for hard-coded implementations, as they can automatically evaluate gradients, thereby facilitating a more straightforward implementation process and enhancing the extensibility to newer models and optimization techniques with ease. In contrast, EM-based methods require hard-coding of update steps for the available models.

To summarize, \texttt{Mixture-Models} library offers three primary advantages over existing Expectation-Maximization (EM)-based R packages, providing strong motivation for its adoption in various research contexts.
First is its ability to fit Gaussian Mixture Models (GMMs) to high-dimensional data without imposing any modeling assumptions. This ability to fit models without stringent constraints can lead to improved clustering performance, an area where traditional EM-based inference tends to struggle due to its limitations in scaling to high-dimensional data.
Second,
the \texttt{Mixture-Models} library also incorporates second-order optimization routines such as Newton-CG. These routines represent a substantial improvement over the first-order EM, with the capacity to expedite convergence, thereby facilitating faster computational processes.
Lastly, the comprehensive nature of this package obviates the necessity to alternate between R and Python to fit various types of mixture models, such as Parsimonious Gaussian Mixture Models (PGMMs), MClust or Mixture of Factor Analyzers (MFAs). All the requisite mixture models are consolidated in this single package, providing a uniform syntax, thereby streamlining the user experience and facilitating ease of implementation.

Since the focus of this paper is introducing the \texttt{Mixture-Models} library, we concentrate on the implementation details and how it compares with existing implementations.
To keep the main text accessible, we have briefly discussed the background in Section 2 and deferred the in-depth technical details to the appendix, with pointers provided at relevant places. The rest of the paper is organized as follows: Section 3 provides an overview and design of the library, Section 4 provides a comparative analysis of \texttt{Mixture-Models} library with \texttt{scikit-learn}'s implemenation, and Section 4 concludes with directions for future development.

\section{Background}
Gaussian Mixture Models (GMMs) have long been a fundamental tool in model based clustering. A GMM is a probabilistic model that assumes the data is generated from a mixture of several Gaussian distributions, each with its own parameters; refer to Appendix \ref{app:intro} for more details. This approach allows for a more flexible representation of data distributions than a single Gaussian, making GMMs particularly useful in a variety of applications ranging from clustering and density estimation to pattern recognition.
Frequentist inference in GMMs is primarily achieved through Maximum Likelihood Estimation (MLE). The goal of MLE in the context of GMMs is to find the set of parameters (means, variances, and mixture coefficients) that maximize the likelihood of the observed data. This process involves estimating the parameters such that the probability of the observed data under the assumed model is maximized. Appendix \ref{app:inference_in_MM} contains a brief overview on various approaches for inferring the parameters of GMM.
A notable challenge in using MLE for GMMs is that the likelihood function is known to be degenerate i.e. it is possible to trivially increase the likelihood by fitting a single Gaussian onto a single point and reducing its variance. This degeneracy implies that there are multiple sets of parameters that can yield the same maximum likelihood, making the optimization landscape complex and posing challenges in finding a unique, global solution. Appendix \ref{app:degeneracy} has more details on degeneracy and how it is tackled.

It should be noted that traditionally gradient descent (GD) based approaches were not the preferred choice for frequentist inference of mixture models \footnote{We use the term Gradient {\it Descent} for maximization also, assuming appropriate change of sign during optimization.}. This is due to the fact that parameter estimation in mixture models using GD is a constrained optimization problem e.g. the sum of mixture weights should add up to 1 and covariance estimates should be positive definite. It is possible to convert these constrained optimization problems into unconstrained optimization problems through suitable reparametrizations \citep{kasa2022improved, kasa2023avoiding}. Further through the availability of Automatic Differentiation (AD) tools, it is possible to compute the gradients to evaluate machine precision without the need for getting their expressions first.  For a beginner friendly introduction on AD and how it can be used for inference in GMMs, refer to Appendix D of \citet{kasa2023avoiding}. Thus, using AD-based tools and certain reparametrizations, it is possible to do inference on GMMs without the need for explicitly computing the expressions first. Appendix \ref{app_sec:reparametrizations} gives a detailed overview of the various reparametrizations used in this library and a pseudo-code for fitting GMM using the AD-GD approach. 

One of the reasons why Expectation Maximization (EM) has been the practitioners' favorite is because
EM elegantly solves the constrained optimization problem without need for any reparametrizations. Further unlike several GD based methods, EM, a quasi-super linear method, guarantees a monotonic increase in likelihood without any need for finetuning for hyperparameters \citep{jordanEMasGD}. However, the expressions for iterative updates need to hand-computed ahead for newer models; nevertheless, it is more than 50x faster than using AD-GD approach as the expressions are precomputed (Appendix J, \citep{kasa2023avoiding}).  Appendix \ref{app:reasons_for_EM} contains more details on how EM solves the constrained optimization problem by design.

High-dimensionality (HD) refers to the case where the number of datapoints is of the same order or greater than the number of features. High-dimensionality, in the context of fitting GMMs, leads to several interesting paradigms such as overparametrization, blessing of dimensionality, etc. There have been several constrained variants of GMM proposed to fit on HD data - refer to Appendix \ref{app:hdmodels}. One of the reasons for proposing the constrained variants is because it is not possible to do inference of an unconstrained vanilla GMM using EM in HD settings as the covariance estimates will not be positive definite - refer to Appendix \ref{app:EMinHD}. Thus, preemptively imposing additional constraints/structures to vanilla GMM reduces the overparametrization and enables inference through EM. 
Note that AD-GD approach discussed above does not face this issue and can naturally scale to high-dimensional data without the need for constraints. In general, imposing pre-determined constraints on means or covariance matrices are discouraged and a more data-dependent discovery of cluster structure is recommended, e.g., by \citep{zhou2009penalized,fop2019model}.

\section{Implementation Details: Overview and Design}

In our work, we have constructed a foundational \texttt{MixtureModels} class, from which we have further developed individual subclasses representing Gaussian Mixture Models (GMM), Parsimonious Gaussian Mixture Models (PGMM), Mixture of Factor Analyzers (MFA), MClust, etc. These subclasses have been carefully designed by exploiting the intrinsic hierarchical structure inherent to these mixture models. This approach underscores the modularity and extensibility of our implementation, facilitating future adaptations and expansions.

\begin{figure}[htbp]
  \centering
  \includegraphics[width=\textwidth]{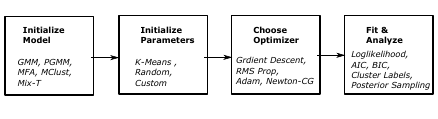}
  \caption{Workflow for using \texttt{Mixture-Models} library}
  \label{fig:workflow}
\end{figure}

A typical API call in our package works as follows:
 \begin{enumerate}
            \item Select model for data: $\mathtt{model = Mclust(data,\: constraint=\texttt{"}VVV\texttt{"})}$
            \item Initialize model parameters: $\mathtt{model.init\_params(num\_components=5)}$
            \item Fit the model: $\mathtt{model.fit(init\_params,\:\texttt{"}grad\_descent\texttt{"},\:\ldots)}$
            \item Export parameters, cluster assignments etc.
\end{enumerate}
Figure \ref{fig:workflow} shows a summary of the different options available at each step of inference. Detailed illustrative examples on how to fit various mixture models and how to access various features of the library can be found here \footnote{\url{https://github.com/kasakh/Mixture-Models/tree/master/Mixture_Models/Examples}}.

On top of developing the core functionality of the \texttt{Mixture-Models} package, we have also  written a test suite to check and ensure that our entire workflow -- from the generation of simulated datasets, to the initialization of model parameters, execution trace of the optimization routine, as well as intermediate and final parameter estimates -- is able to correctly reproduce results of experimental runs. Appendix \ref{app:data_simulation} has more details on various data simulation options present in our library.

\section{Simulations}

We conducted a comprehensive benchmark study to evaluate the efficacy of various optimization techniques, including vanilla Gradient Descent, Adam, and Newton-CG, against the well-established EM algorithm as implemented in \texttt{scikit-learn} library. This evaluation is conducted under a diverse range of conditions, characterized by varying the number of data points (n), feature dimensions (p), and mixture components (K). Specifically, the means of each GMM component are generated using \texttt{np.random.rand(K, p) * scale}, providing a random but scaled representation of the feature space. Additionally, the covariance matrices are derived from \texttt{np.random.randn(K,p,p) / np.sqrt(p)}, ensuring a realistic level of complexity and variance within each simulated dataset. The values of $n,p \; \text{and } K$ are varied as follows - $n = \{32,64,128,256,512\}$, $p = \{2,\dots,9\}$, and $K = \{2,3,4,5\}$. For each value of $(n,p,K)$, we simulate 10 different GMMs using random seeds. 

For each of these datasets, we fit a GMM using four different optimization routines - Gradient Descent (GD), Adam, Newton (implemented in \texttt{Mixture-Models})   and 
EM (implemented in \texttt{scikit-learn}). For GD, the learning rate set to be $3e-4$ for all experiments. For Adam, the learning rate and momentum are set to be $3e-4$ and $0.9$ respectively. All the methods are run till maximum iterations of 1000 or convergence (tolerance is set to be $1e-6$) whichever is met earliest.

We set $\texttt{scale}=5$ to ensure the components are well separated. This extensive and systematic approach allows us to draw meaningful conclusions about the performance and suitability of each optimization technique across a variety of GMM configurations, offering valuable insights into their practical applications in complex, high-dimensional data environments.

\begin{figure}[H]
  \centering
  \includegraphics[width=0.9\textwidth]{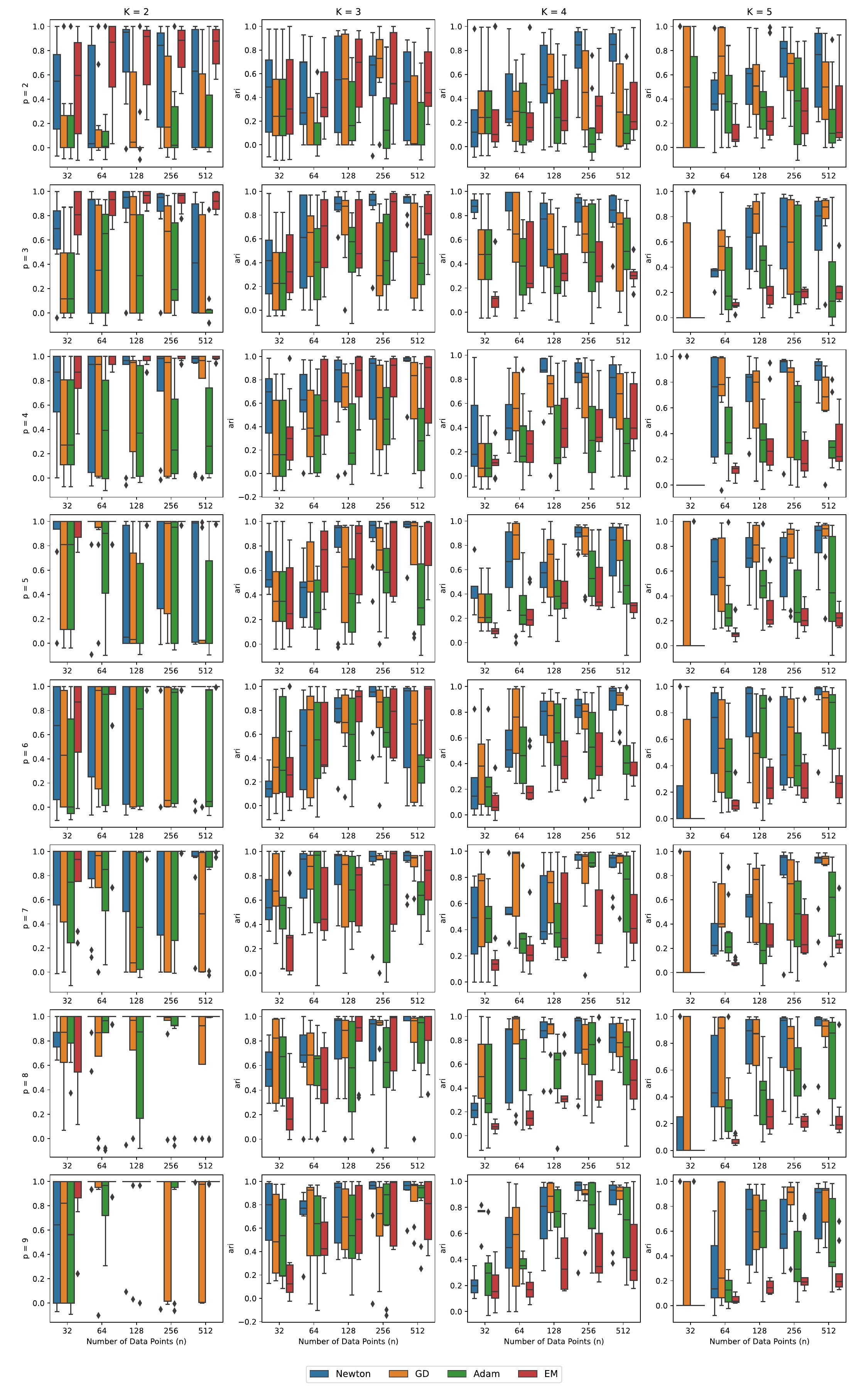}
  \caption{Comparing the performance of Gradient Descent (GD), Adam, Newton-CG and EM by varying $n$, $p$ and $K$}. 
  \label{fig:comparison}
\end{figure}

The results are plotted in figure \ref{fig:comparison} where in each subplot contains boxplots for a given $(p,K)$ with varying $n$. First, we note that as the number of clusters $K$ increases, gradient based methods as GD, Adam and Newton-CG outperform EM. Next, we note that this improvement is further accentuated with increasing dimensionality $p$. Interestingly, EM exhibits a relatively lower variability in performance across all the settings as compared to the gradient based methods.

\section{Conclusion and Future Work}

The \texttt{Mixture-Models} is the first python based library to present an one-stop solution to the challenges of fitting GMM and its variants, especially in high-dimensional settings. Its emphasis on user-friendliness, modularity, and advanced optimization techniques, along with comprehensive documentation make it an invaluable tool for both academic and industry practitioners. Under MIT license, the library is freely available, encouraging its adoption in the wider research community. In the current version, we implemented various GMM based mixture models whose inference is carried out using gradient based methods using \texttt{autograd} library in Python. In the future, we aim to enhance the library by ensuring compatibility with PyTorch, thereby bringing mixture models more accessibly to the deep learning community.

\bibliography{references}

\newpage

\appendix

\section{Gaussian Mixture Models}
\label{app:intro}
Let $f(\boldsymbol{x} ; \boldsymbol{\btheta})$ be the density of a $K$-component mixture model.
Let $f_k$ denote the $k^{\rm th}$ component density with parameters  $\btheta_k$ and weight  $\pi_k$.
The density of the mixture model is given by
$$f(\boldsymbol{x} ; \boldsymbol{\btheta}) =\sum_{k=1}^{K} \pi_{k} f_{k}\left(\boldsymbol{x} ; \boldsymbol{\btheta}_{k}\right), $$
where  $\sum_{k=1}^{K} \pi_k = 1$ and $\pi_k \ge 0$ for  $k = 1,\ldots, K$ and
$\btheta$ %
denotes the complete set of parameters of the model. %
In a GMM, each individual component $f_k$ is modeled using a multivariate Gaussian distribution $\mathcal{N}(\bmu_k, \bSigma_k)$ where $\bmu_k$ and $\bSigma_k$ are its mean and covariance respectively.

Given $n$ independent and identically distributed  $(iid)$
instances of $p$-dimensional data, $[x_{ij}]_{n \times p} $
where index
$i$ is used for observation, and $j$ is used for dimension.
Maximum Likelihood Estimation (MLE) 
aims to find parameter estimates
$\hat \btheta $ from the overall parameter space $\bTheta$ of $f(\btheta)$ such that probability of observing the data %
samples $\bx_1, \dots, \bx_n$ from this family 
is maximized, i.e.,
$\hat{\btheta}=\argmax_{\btheta \in \bTheta} \mathcal{L}(\btheta)$, where, 
$\mathcal{L(\btheta)} = \frac{1}{n}\sum_i \log f(\bx_i;\btheta) $ is the loglikelihood.

\section{Inference in Mixture Models}
\label{app:inference_in_MM}
{\bf Expectation Maximization (EM)},
a widely used method for MLE of GMM, proposed by \cite{dempster1977maximum}, alternates between optimizing the component parameters (means \& covariances) and component weights. \cite{wu1983convergence} has shown that when the likelihood is unimodal, the algorithm converges to the global optimal. However, this is not the case with mixture models whose likelihood is multimodal.  \cite{balakrishnan2017statistical} provide statistical guarantees that EM converges to local optimum which close to the global optimum under suitable initialization {\color{black} for GMMs}.  
{\color{black}Maximizing the likelihood for GMMs is NP-hard \citep{kannan2005spectral}}.

Although other methods have been proposed, 
maximizing likelihood is by far the most popular method of inference in mixture models.   Several algorithms based on pairwise distances, spectral methods, method of moments, Fischer's Linear Discriminant Analysis (LDA), subspace clustering methods, variable selection and Mixed Integer Optimization (MIO) methods have been proposed. 
Pairwise distance based algorithms, which involve projecting the datapoints onto random directions such that the intra-cluster distance is reduced whereas the intercluster distance is maximized, have been studied in  \citep{dasgupta1999learning, dasgupta2000two, sanjeev2001learning}. A drawback of these algorithms is that they require the mean separation to increase with dimensionality. The spectral methods studied in \cite{vempala2004spectral, kannan2005spectral, brubaker2008isotropic} overcame this problem of increasing mean separation with dimension, by choosing a subspace based on large principal components and projecting on this subspace. Nevertheless, 
spectral clustering algorithms also need the Gaussian componenets in the mixture to be separated by at least a constant times the standard deviation. \cite{kalai2012disentangling} overcame this requirement of mean separation by using method of moments. \cite{bandi2019learning} proposed an MIO based algorithm that relies that on constructing a model by minimizing the discrepancy between the empirical CDF and model CDF. The discrepancy measures studided were Kolmogrov Smirnoff distance and Total Variation distance. While the runtime of this proposed algorithm is significantly higher compared to EM, it offers better out-of-sample accuracy even when the mean-separation is low. \cite{pan2007penalized,guo2010pairwise,raftery2006variable} study simultaneous variable selection and clustering with applications in bioinformatics. A variable is considered irrelevant if its mean is same across all the components. In the later sections we study a complementary scenario where all the components have the same mean.  \cite{azizyan2013minimax} provides the theoretical justification for these variable selection based clustering methods. \cite{azizyan2015efficient} points out the cases where feature selection based methods can fail. In order to come this drawback, they propose an LDA based method for the case of two component of GMM. 
EM and its variants remain the most widely used methods to obtain ML estimates of GMM for clustering.

\section{Degeneracy}
\label{app:degeneracy}

In vanilla GMMs, the idea of a true MLE does not exist due to degeneracy - this is because one could always assign a single datapoint to a cluster, the variance of this component can be made as small possible while arbitrarily increasing the likelihood as large as possible. \citep{day1969estimating,ingrassia2004likelihood,Chen2009,ingrassia2011degeneracy}. A consequence of the unboundness of the likelihood in vanilla GMMs is the \textit{spurious} clustering solutions,
which are local maximizers of the likelihood function but lack real-life interpretability and hence do not provide a good clustering of the data. Degeneracy can be tackled by imposing restrictions on the size of the covariance estimates such that they do not become singular, e.g. \cite{ciuperca2003penalized}, or by imposing constraints such that the covariance is same across all the components. We refer to Appendix B of \cite{kasa2023avoiding} for more details on degeneracy and spurious solutions.

\section{Reasons for preferring EM}
\label{app:reasons_for_EM}

Typically Expectation Maximization (EM) is used to solve the problem of maximizing the likelihood of $\mathcal{L}(X|\theta)$ where $X$ is the observed data and $\theta$ is the parameter vector. Any introductory material on EM starts off by showing how EM simplifies the maximization in the case of Gaussian Mixture Models (E.g: Chapter 9, PRML, C. Bishop). EM and its variants are widely used in the problems of latent/missing/unobserved data. There are a few reasons why EM has been so popular. Some of them are discussed below:

\begin{itemize}
	\item Solving the $\text{argmax}_{\theta}\mathcal{L}(X|\theta)$ problem is typically intractable if one tries to evaluate the gradients and set them to zero. Using EM, one tries to reformulate the above problem using its lower bound and tries to maximize the lower bound. It is expected that maximizing this lower bound is tractable and easier. 
	
	\item In the case of GMMs,  the parameter vector contains the covariance matrices $\Sigma_g$ for each component $g$. 
	\item \textit{Problem 1:} Evaluating derivatives wrt to matrices is inconvenient i.e. $\frac{\partial L }{\partial \Sigma_g}$ is almost intractable. Using EM, we simplify the problem into a sequence of simpler optimization problems where the gradients are easy to obtain and immediately set to zero, i.e. no need for any gradient descent approach.  
	
	\item \textit{Problem 2:} Moreover, without EM, maximizing $\mathcal{L}(X|\theta)$ with $\Sigma_g$ would require additional constraints on the positive semi definiteness (PSD) of the estimate $\hat{\mathcal{L}(X|\theta)}$ we are trying to obtain. However, EM eliminates the need for any such constraints by nature of its formulation.
	
	\item \textit{Problem 3:} To ensure that mixture weights remain positive and add up to one, typically a Lagrange multiplier is used.

\end{itemize}

\section{Reparametrizations}
\label{app_sec:reparametrizations}
\begin{itemize}
\item To ensure Positive Definiteness (PD) of the estimated covariance matrices, 
we compute the gradients with respect to $\bV_k$, where $\bSigma_k = \bV_k \bV^{T}_k$. We first initialize the values of $\bV_k$ as identity matrices. Thereafter, updated values of $\bV_k$ can be computed using gradient descent, i.e., $\bV_k := \bV_k + \epsilon \times \frac{\partial \mathcal{L}}{\partial \bV_k}$. Here, $\epsilon$ is the learning rate. %
If the gradients are evaluated with respect to $\bSigma_k$ directly, there is no guarantee that updated $\bSigma_k = \bSigma_k + \epsilon \times \frac{\partial \mathcal{L}}{\partial \bSigma_k}$ will still remain PD. However, if gradients are evaluated with respect to $\bV_k$, no matter what the updated matrix $\bV_k$ is, by construction $\bSigma_k$ always remains PD. Alternatively, one could use a Cholesky decomposition for $\bSigma_k$ \citep{salakhutdinov2003optimization}.

\item To ensure that the mixture proportions of components lie in $(0,1)$ and add up to one, we use the logsumexp trick \citep{murphy2012machine}. We start with unbounded $\alpha_k$'s as the log-proportions, i.e.,  $\log \pi_k = \alpha_k - {\log(\sum_{k^{'}} e^{\alpha_{k^{'} }})}$.  Note that, we need not impose any constraints on $\alpha_k$ as final computation of $\pi_k$ automatically leads to normalization, because $\pi_k = \frac{\alpha_k}{\sum_{k^{'} } e^{\alpha_{k^{'} } }}$. Therefore, we can update $\alpha_k := \alpha_k + \epsilon \times \frac{\partial L}{\partial \alpha_k}$ without any further need for Lagrange multipliers.

\item In models such as MClust, there is an additional constraint that the orientation matrix estimate is orthogonal. This constrained optimization problem is converted into unconstrained problem using Caley Transform \citep{cayley1846quelques} as follows. Given an $p \times p$ matrix A, the orthogonal matrix $O$ can be written as $O = (I+Z)^{-1}(I-Z)$ where $Z = (A - A^{T})/2$ is a skew-symmetric matrix. Instead of performing gradient descent on $O$, we peform gradient updates on $A$ and update $O$ using the Cayley Transform. 

\item In mixture of T distributions, the degrees of freedom estimate $\nu$ needs to be ensured positive. This can be ensured by the transform $\nu = e^{\nu^{'}}$ and doing gradient updates on $\nu^{'}$.

\end{itemize}

Below we give the algorithm for fitting an unconstrained GMM. Using the above reparametrizations, the same can be extended for other mixture models such MClust, PGMM, etc.

\begin{algorithm}[h!]
		\caption{AD-GD inference of GMM}
		\label{ALGO:SIA_combined}
		
		\textbf{Input: }
		Data: $n \times p$ dimensional matrix, 
		number of clusters $K$, 
		learning rate $\epsilon$, 
		convergence tolerance $\gamma$.
		
		\textbf{Initialize} at $t=0$: 
		$\hat{\bmu}^{0}_k, \hat{\alpha}_k$'s using K-Means or random initialization; 
		$\hat{\bU}^{0}_k$ (for GMM) 
		as identity matrices.
		
		\textbf{REPEAT:} At every iteration $t+1$:

			\begin{minipage}{\linewidth}
				\begin{align*}
				\hat{\alpha}^{t+1}_k := \hat{\alpha}^{t}_k + \epsilon \frac{\partial \mathcal{L} }{\partial \alpha_k} ; \; 
    \hat{\pi}^{t+1}_k  := \frac{ e^{\hat{\alpha}^{t+1}_k}}{\sum_{k^{'}} e^{\hat{\alpha}^{t+1}_{k^{'} }}} ; \;  \hat{\bmu}^{t+1}_k := \hat{\bmu}^{t+1}_k + \epsilon  \frac{\partial \mathcal{L}}{\partial \bmu_k} ; \;
				\end{align*}
			\end{minipage}

		\resizebox{0.9\linewidth}{!}{
			\begin{minipage}{\linewidth}
				\begin{align*}
				\hat{\bV}^{t+1}_k := \hat{\bV}^{t}_k + \epsilon \frac{\partial  \mathcal{L}}{\partial \bV_k} ; \;
				\hat{\bSigma}^{t+1}_k := \hat{\bV}^{t+1}_k \hat{\bV}^{{t+1}^T}_k 
				\end{align*}
			\end{minipage}
		}
		
		\textbf{UNTIL:} convergence criterion $|\mathcal{L}^{t+1} - \mathcal{L}^{t} | < \gamma $ is met
		
	\end{algorithm}

\section{High Dimensional Data} \label{app:hdmodels}

In this section, we briefly introduce High-Dimensional (HD) data, discuss some issues related to HD data and previous literature related to tackling HD data in the context of GMMs. High-dimensionality refers to the case where the number of datapoints $(n)$ is less than or equal to the number of features $(p)$. High-dimensionality leads to several interesting paradigms such as blessing of dimensionality, parsimony, etc. in model based clustering as described below:

\subsection{Blessing of Dimensionality}
When $n \le  p$ our model is overparameterized and it is not possible to reliably estimate all the parameters. For the $K$-component $p$-dimensional Gaussian Mixture Model, there $K-1$, $K\times p$ and $K\times p \times (p-1)$ parameters to be estimated for the mixing weights, means and covariance matrices respectively. As it is evident, the number of parameters to be estimated grows quadratically in $p$, hence the number of datapoints need to grow at least quadratically for reliable estimation. 

On the brighter side, the sparsity of data in high-dimensions was shown to be a blessing with regards to Clustering \citep{bouveyron2014model}. \cite{huberrobust} suggested an experimented for drawing samples from $p$-dimensional hypersphere for radius 1 with uniform probability. The probability that sample is drawn from the shell between a radius r and 1 is given by $1 - r^p$. For example, consider a 10-dimensional hypersphere, the probability that a distribution lies in the shell between 0.8 and 1 is approximately 90\%. Hence, it may be beneficial to explore only the low-dimensional subspaces in which the datapoints may lie. The subspace clustering methods rely on this low-dimensional subspaces to model the data. 

\begin{figure}[H]
	\centering
	\includegraphics[width=8cm]{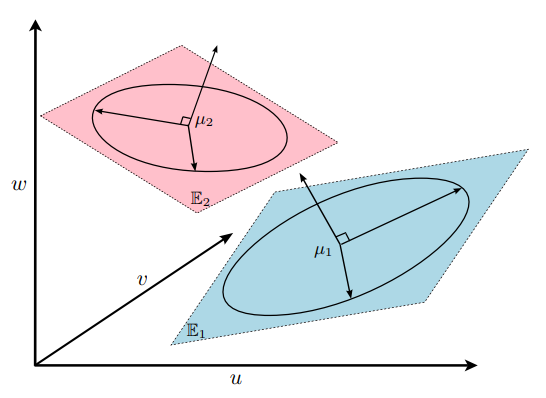}
	\caption{Illustration of Subspace Clustering(taken from \cite{bouveyron2007high})}
	\label{subspace}
\end{figure}

\begin{figure}[H]
	\centering
	\includegraphics[width=12cm]{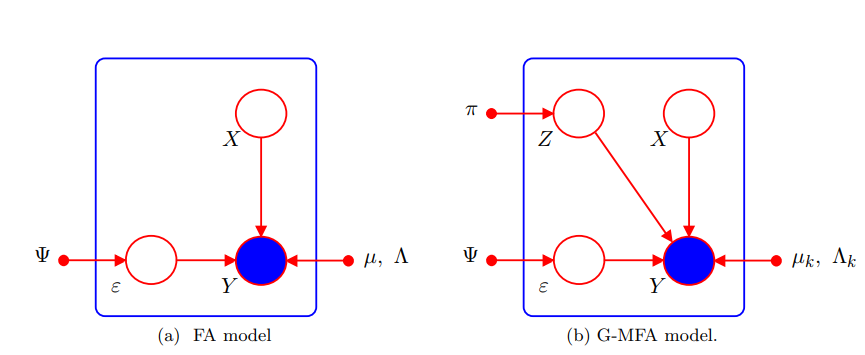}
	\caption{Comparison of FA vs MFA(taken from \cite{bouveyron2014model})}
	\label{favsmfa}
\end{figure}

One of the first subspace model based clustering is Mixture of Factor Analyzers model \citep{ghahramani1996algorithm, mclachlan2003modelling}. The model assumes that that the observed $p$-dimensional data $\bX$ is generated by the $k$-th cluster is related to local latent variable $\bY$ which exists in a subspace $q$ and is related to the observed variable $\bY$ as follows

\begin{align}
\bX_{ | Z=k}=\bLambda_{k} \bY+\bmu_{k}+\varepsilon
\end{align}

where, $\bLambda_k$ is a $p\times q$ loading matrix of the $k$-th component, and $\bmu_k$ is the mean of the $k$-th component. MFA can be considered as the extension of Factor Analyzers (FA) to mixture models. The comparison between the two models is given in figure \ref{favsmfa}

\subsection{Dimensionality Reduction}
It is important to notice the difference between dimensionality reduction techniques such as Principal Component Analysis (PCA)  vs Latent Factor models such FA or MFA. PCA and its variants try to capture the direction of maximum variance in the form of eigenvalues and eigenvectors. FA and its variants assume that there is a latent low-dimensional factor that generates the data which is observed with some error. Practically, FA model reduces the dimensions while keeping the covariance structure the same \citep{bouveyron2014model}. In particular, several researchers \citep{pan2007penalized, yeung2001principal, raftery2003discussion}  have cautioned against using PCA as dimensionality reduction technique before using clustering. This are two reasons for this:
\begin{itemize}
	\item \cite{chang1983using} has shown theoretically and empirically that identifying the larger eigenvectors do not contain the discriminating information compared to others.
	\item PCA lacks the interpretability which is usually present in FA models. Moreover, as discussed in previous section, it is easier to discriminate in high dimensions as compared to low-dimensions, assuming one has an algorithm that works in high dimensions (typically, MFA models do well in HD)  
\end{itemize}

\subsection{Parsimonious models}
The main problem of high-dimensional modeling is overparameterization and Parsimonious models such as PGMMs \cite{McNi:Pgmm:2011, mcnicholas08,mcnicholas10a} overcome this by imposing a constraint on the covariance matrices $\bSigma_K$ of the components. With $p$-dimensional data and with $K$ components in a mixture there are exactly $(K-1)+Kp+Kp(p+1)/2$ free parameters. Unless $p$ is very small, most of these parameters are in the component covariance matrices $\matsig_1,\ldots,\matsig_K$. To reduce the computational cost of the estimation, special families of covariance structures have been introduced that impose constraints upon the constituent parts of the decomposition of $\matsig_k$. 
The Parsimonious Gaussian Mixture family, or PGMM  \citep{mcnicholas08} are a family wherein 
the covariance structure is assumed to be of the form
$\bSigma_k=\bLambda_k\bLambda_k'+\mathbf{\Omega}_k$, where $\mathbf{\Omega}_k$ 
is a diagonal matrix of white noise, %
$\Lambda_k$ is a $p$ x $q$ matrix of factor loadings and $q$ is the number of latent factors. Generally $q < p$. The loading and noise terms can be constrained to be equal or unequal across groups to give a collection of eight parsimonious covariance structures shown in table \ref{PGMM}. The number of parameters in these eight families range from $pq-q(q-1)/2+1$ to $K(pq-q(q-1)/2+p)$. See  \cite{mcnicholas08} for more details. 

\begin{table}[t]
	\footnotesize
	\centering
	{
		\begin{tabular}{lr}
			\hline
			Model  & Covariance Parameters\\
			\hline
		CCC  & $[pq-q(q-1)/2]+1$\\
		CCU  & $[pq-q(q-1)/2]+p$\\
		CUC  & $[pq-q(q-1)/2]+K$\\
		CUU  & $[pq-q(q-1)/2]+Kp$\\
		UCC  & $K[pq-q(q-1)/2]+1$\\
		UCU  & $K[pq-q(q-1)/2]+p$\\
		UUC  & $K[pq-q(q-1)/2]+K$\\
		UUU  & $K\{pq-q(q-1)/2\}+Kp$\\
			\hline
		\end{tabular}
	}
	\caption{Parsimonious Gaussian Mixture Models%
	}\label{PGMM}
\end{table}

\cite{bouveyron2007high} developed HDDC algorithm, that uses a combination of subspace clustering and parsimonious modeling for GMMs. The key idea is to decompose the variance $\bSigma_k$ of $k$-th component using SVD as follows:

\begin{align}
\Sigma_{k}=Q_{k} \Lambda_{k} Q_{k}^{t}
\end{align}

where $\Lambda_{k}=\operatorname{diag}\left(a_{k 1}, \ldots, a_{k d_{k}}, b_{k}, \ldots, b_{k}\right)$ is the diagonal matrix of eigenvalues and $Q_k$ is the orthogonal matrix of corresponding eigenvectors. As can be seen from $\Lambda_k$ the magnitude of the $p-d_k$ noise terms is the same value $b_k$. Essentially, the data is assumed to be in a $d_k$-dimensional subspace and the noise is assumed to be orthogonal to this subspace. The authors propose 16 different models based on varying or constrained these noise and subspaces parameters across all the dimensions. Unlike PGMMs where inference is carried out using AECM, inference is carried out using EM.

\section{EM in high dimensions}\label{app:EMinHD}

In this section, we discuss the problems related to EM algorithm in High Dimensional setting and how those problems were overcome.
\begin{enumerate}
	\item In the Expectation step (E-step), one need to invert a weighted sample covariance matrix i.e. 
	\begin{align}
	\gamma\left(z_{n k}\right)=\frac{\pi_{k} \mathcal{N}\left(\mathbf{x}_{n} | \boldsymbol{\mu}_{k}, \mathbf{\Sigma}_{k}\right)}{\sum_{j=1}^{K} \pi_{j} \mathcal{N}\left(\mathbf{x}_{n} | \boldsymbol{\mu}_{j}, \mathbf{\Sigma}_{j}\right)}
	\end{align}
	In HD setting, this matrix does not usually have a full rank. Intuitively, it is because if $\bX$ is the data matrix with $p$-rows and $n$-columns and $p>n$, then rank of $p\times p$ covariance matrix  $\bX\bX^T$ is $\min(n,p)$ which is $n$. 
	\item There have been several attempts to tackle this problem by using constraints on the structure of covariance matrix such as Mixture of Factor Analyzers (MFA) \citep{ghahramani1996algorithm} and Parsimonious GMMs \citep{mcnicholas08} - both of which use AECM algorithm for inference.
 {\color{black}{Very Recently a new algorithm for high-dimensional GMMs has been proposed - CHIME \citep{cai2019chime} -  that uses a modified E-step to tackle the low-rank covariance matrix. It specifically uses a modified Fisher's Linear Discriminant function $\bbeta$ to allocate each datapoint to one of the clusters.
			
			\begin{align}
			\hat{\boldsymbol{\beta}}^{(t+1)}=\underset{\beta \in \mathbb{R}^{p}}{\arg \min }\left\{\frac{1}{2} \boldsymbol{\beta}^{\top} \hat{\Sigma}^{(t+1)} \boldsymbol{\beta}-\boldsymbol{\beta}^{\top}\left(\hat{\boldsymbol{\mu}}_{k}^{(t+1)}-\hat{\boldsymbol{\mu}}_{1}^{(t+1)}\right)+\lambda_{n}^{(t+1)}\|\boldsymbol{\beta}\|_{1}\right\}
			\end{align}
			
			As evident from the above equation, there is no need to compute the inverse of the covairance matrix $\hat{\bSigma}$ . Moreover, for high-dimensional cases, a sparsity on the discriminant function $\bbeta$ is assumed. As pointed out in \citep{park2009singularity, amari2006singularities}, to avoid the milnor attractors, they give basins of concentration for initialization which ensures convergence.  One of the drawbacks of this method is that it assumes equi-covariance $\hat{\bSigma}$  on all the clusters. More importantly, this modified E-step requires solving a Quadractic Optimization problem for the K-1 clusters}}
\end{enumerate}

\section{Data Simulation in \texttt{Mixture-Models} library}
\label{app:data_simulation}

For ease of testing, we have bundled several popular classification datasets into our package, including the \texttt{iris} flower dataset~\citep{anderson1936species,fisher1936use}, the Italian \texttt{wine} dataset~\citep{forina2008parvus}, as well as the \texttt{Khan} gene dataset~\citep{khan2001classification}, with the latter two datasets copied from the UCI Machine Learning Repository~\cite{Dua:2019}. We have also written a sampling routine for generating simulated datasets from a mixture distribution $f(X|\theta)=\prod_{i=1}^n\sum_{k=1}^K\pi_kf_k(x_i|\psi_k)$, allowing for various characteristics according to the factors in our experimental design:
\begin{itemize}
    \item Number of observations $n$, data dimension $p$, and number of clusters $K$ (which we collectively refer to as the \emph{problem size variables}).
    \item Presence vs absence of imbalanced classes: Imbalance is a well-known obstacle in data classification~\cite{japkowicz2002class}, and can be viewed as a generalization of the instability risk discussed in Section 2.4 for Gaussian mixture models.
    \item Presence vs absence of uninformative features: For high-dimensional datasets, the approach of \emph{variable selection}~\cite{wang2008variable} aims to identify a small subset of features that are sufficient to explain the clustering structure. We believe it is of interest to investigate if the AD and EM algorithms are robust to the inclusion of uninformative features (that are statistically indistinguishable from random noise).
    \item Full vs constrained specification for the data generating process: This enables us to investigate the effect of model underspecification/overspecification. Additionally, in the analysis of high-dimensional datasets, it is a common working assumption~\cite{bickel2004some,wang2008variable} that the $p$ features are independent and that the clusters have common covariance, in other words: $\Sigma_k=\Sigma=\mathrm{diag}(\sigma_1^2,\dots,\sigma_K^2)$ for each cluster $k=1,\dots,K$. We provide this option, along with an alternate specification of distinct (but still diagonal) covariances $\Sigma_k=\mathrm{diag}(\sigma_{k,1}^2,\dots,\sigma_{k,K}^2)$ for each cluster\footnote{In the terminology of the MCLUST~\cite{fraley2003enhanced} model specifications, these processes can be described as ``EEI'' and ``VVI'' constraints, respectively.}, the latter in order to support comparison with the \texttt{scikit-learn} implementation of EM (which has the option of fitting models of this form).
\end{itemize}

\end{document}